# Unexpected effect of Ru-substitution in lightly doped manganites

**Lorenzo Malavasi,**[*a] **M. Cristina Mozzati**[b]**, Cristina Tealdi**[a]**, M. Rosa Pascarelli**[a]**, Carlo B. Azzoni**[b] **and Giorgio Flor**[a]

[a] *Department of Physical Chemistry "M. Rolla", University of Pavia, 27100 Pavia, Italy. Fax: +39-(0)382-507575; Tel: +39-(0)382-507921; *E-mail: lorenzo.malavasi@unipv.it*
[b] *Department of Physics "A. Volta", INFM, University of Pavia, 27100 Pavia, Italy.*

*This submission was created using the RSC ChemComm Template (DO NOT DELETE THIS TEXT)*
*(LINE INCLUDED FOR SPACING ONLY - DO NOT DELETE THIS TEXT)*

**In this Communication we report about the unexpected effect of ruthenium doping in sodium ligthly-doped manganites. This effect seems to be in contrast with the usual model applied to describe the effect of this magnetic ion into the manganite structure. We propose a possible compensation mechanism which seems also able to describe other peculiar features encountered in these materials.**

Recently it was reported by several independent groups that the ruthenium doping of perovskite manganites is an effective route to enhance their magnetic and transport properties[1-3]. This is particularly true for the charge-ordered (CO) manganites where the various possible types of antiferromagnetic (AF) orderings are removed by Ru substitution on the Mn-site. However, on optimally doped manganites, *i.e.* for Mn(IV) values around 30-33%, the available literature papers report slightly reduction of Curie temperatures ($T_C$) and semiconducting to metal transitions ($T_{SM}$).

The peculiar effect of Ru-doping, with respect to other 3d or 4d transition metals, seems to be directly correlated to its ability to preserve and participate to the double-exchange (DE) mechanism[4]. This in turn seems to be due to its electronic configuration and to the fact that the ruthenium ions can adopt two different oxidation states, +4 and +5, as also evidenced experimentally in CO and optimally doped manganites[4-5]. In both oxidation states Ru has vacant $e_g$ orbitals (low spin systems) and so a possible electron transfer can occur between Mn and Ru ions.

In attempts to deepen the role of Ru-doping on Mn site we started to study the $La_{1-x}Na_xMn_{1-y}Ru_yO_{3+\delta}$ system (LNMRO) also for monovalent dopings far from the optimal one. The samples with $x$ = 0.05 and 0.15 and $y$ = 0, 0.05 and 0.15, were prepared by solid state reaction at 960°C starting from high purity metal oxides and/or carbonates. After synthesis two batches of the as-prepared samples were annealed in pure oxygen and pure argon, respectively, for 72 h at 900°C and quenched to room temperature. The samples were thoroughly characterized by means of X-ray diffraction, electron microprobe analysis, thermal analysis, resistivity measurements from 300 to 10 K without applied magnetic field and with a field of 1 and 7 T, by magnetic susceptibility and electron paramagnetic resonance measurements. In this communication we will report only part of the whole set of results we gained from this research. The overall data will be published later in a complete work.

Figure 1 reports the zero field cooled (ZFC) and field cooled (FC) *M vs T* curves at 100 G for the three samples (oxygen annealed) with $x$ = 0.05 and $y$ = 0 (circles), 0.05 (squares) and 0.15 (triangles). As can be appreciated, the $T_C$ (taken at the inflection point of the FC curve) is spectacularly more than doubled from 117 K for the Ru-free sample (from now N5R0) to 251 K for the $La_{0.95}Na_{0.05}Mn_{0.95}Ru_{0.05}O_{3+\delta}$ (called N5R5 in the rest of the paper). For $La_{0.95}Na_{0.05}Mn_{0.85}Ru_{0.15}O_{3+\delta}$ (called N5R15) the $T_C$ then decreases to 212 K.
For a Na-doping of 0.15 (N15 series) the $T_C$ shows a small reduction from N15R0 to N15R5 (from 305 to 301 K) and a more pronuced fall to 172 K when the Ru-doping is 0.15 (N15R15 sample).

Figure 2 reports the resistivity ($\rho$) *vs T* behaviour for the N5 series. In particular, the 0, 1 and 7 T curves for the oxygen annealed N5R5 sample are presented in the main plot. In the inset, the logarithm of resistivity for the N5R0 and N5R15 samples are shown. It is clear the presence of two distinct transitions for the N5R5 sample with the first sharp one located around 250 K and the second wider one around 185 K. The latter is not shifted by the magnetic field. The MR value (defined as MR = - [($\rho_{7T}$ - $\rho_{0T}$) / $\rho_{0T}$]) correspondent to the first transition is about 60% (at 260 K).

The presented data show that a small Ru doping on a Na-doped manganite far from the optimally doping can induce an outstanding increase of transition temperatures and the appearance of a metallic state (let's remember that the N5R0 sample is insulating in the whole temperature range explored, see inset of Fig. 2).

This effect can be correlated to the electronic properties of Ru ions which can participate directly in the DE and in a very effective broadening of the conduction band. However, the data for the N5R5 and increasingly for the N5R15 samples seem to indicate that a simple internal oxidation-reduction mechanism, usually suggested for Ru-doped maganites, can not completely explain the present results since the Ru(IV), and even more the Ru(V), should reduce one or two Mn(IV) to Mn(III) (note that in the case of the N5R15 sample this mechanism should create also Mn(II) ions). The lattice constants determined from Rietveld refinement of our patterns (see for example the N5R5 refined pattern deposited as Figure 3 ESI material) and reported in Tabel 1 for the oxygen annealed sample, are a first, even though indirect, indication that this scenario could not be so easy. In fact, passing from N5R0 to N5R5 the cell volume is practically unchanged while it slightly increases going to N5R15. Moreover, from thermogravimetric results (see Figure 4 deposited as ESI material) we had evidence of an increase of oxygen content along with the increase in the Ru-doping. What can be expected is that the substitutional Ru-defect can be partially, or totally, compensated by means of an uptake of oxygen according to:

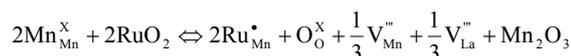

$$2Mn_{Mn}^X + 2RuO_2 \Leftrightarrow 2Ru_{Mn}^\bullet + O_O^X + \frac{1}{3}V_{Mn}''' + \frac{1}{3}V_{La}''' + Mn_2O_3$$

in this case the only immediate effect of ruthenium doping should be the creation of cation vacancies located on the perovskite sites.

This model seems also capable to explain the two main anomalies found in Ru-doped manganites namely the "double bump" characteristic of resistivity curves and the low experimental magnetic moment detected with respect to the theoretical one. As we could assess for LCMO and LNMO systems[6,7], the presence of point defects may cause the appearance of double-peaks in the $\rho$ *vs T* curves. Moreover, the break of interaction paths between magnetic ions caused by the cationic vacancies may explain the unexpected reduction of the magnetic moment. Within the framework of this model, the reduction of $T_C$ detected for the N5R15 member can be nicely accounted for. In fact, even though we introduce, by means of Ru-doping, 0.15 "Zener ions" which should lead to a relevant increase of $T_C$ (to ~ 280 K if we use the "phase diagram" we constructed for the LNMO system[7]), the cation vacancies percentage added should reduce the Curie temperature of about 60-70 K according to[8]. Finally, this increased amount of defects causes the system to became insulating as we have found for Ru-free samples. A first direct proof of the valildity of our model came from preliminary XAS measurements perfomed at the ELETTRA synchrotron in Trieste (Figure 5, ESI material) where the Mn $L_{2,3}$ edge for the N5R0, N5R5 and N5R15 samples are reported: as can be clearly seen in that Figure the shape and position of the three spectra is practically the same, thus confirming that the Ru-doping did not affected the Mn(III)/Mn(IV) ratio as forseen by our model

To conclude, in this paper we report an unexpected effect of Ru-doping on the physical properties of lightly doped LNMO manganites where the effectiveness of this ion in promoting the DE is remarkable. We propose a model to explain our data which, however, could be of applicability also for other systems. Our hypothesis are based on the idea that the ruthenium oxidation state is mainly (IV) (and most of the published papers agree with this scenario and also our preliminary XAS data confirm this feature) and that the substitutional defect created when the ruthenium is introduced in the perovskite structure is not directly compensated by a reduction of the Mn sub-lattice but by a gas-solid equilibrium which involves the uptake of oxygen. This is nicely shown by the likeness of the Mn $L_{2,3}$ edge of the N5 series samples (see the ESI material, in particular Figure 5) which up to now have been characterized by XAS spectroscopy. A more quantitative and complete characterization fo the electronic properties of the LNMRO systems will be published in a separate work.

The Authors acknowledge Dr. Federica Bondino, Dr. Mauro Platè and Prof. Fulvio Parmigiani for performing XAS measurements and the financial support from PRIN project 2002 of Ministry of University of the Italian Government.

**Table 1** Lattice parameters of the $La_{1-x}Na_xMn_{1-y}Ru_yO_{3+\delta}$ system annealed in pure oxygen. For sample names abbreviations see the text.

|  | $a$ (Å) | $b$ (Å) | $c$ (Å) | $V$ (Å$^3$) |
|---|---|---|---|---|
| N5R0 | 5.524 | 5.524 | 13.342 | 58.75 |
| N5R5 | 5.521 | 5.521 | 13.350 | 58.744 |
| N5R15 | 5.526 | 5.526 | 13.345 | 58.82 |
| N15R0 | 5.504 | 5.504 | 13.332 | 58.292 |
| N15R5 | 5.512 | 5.512 | 13.339 | 58.488 |
| N15R15 | 5.518 | 5.518 | 13.345 | 58.644 |

## Notes and references

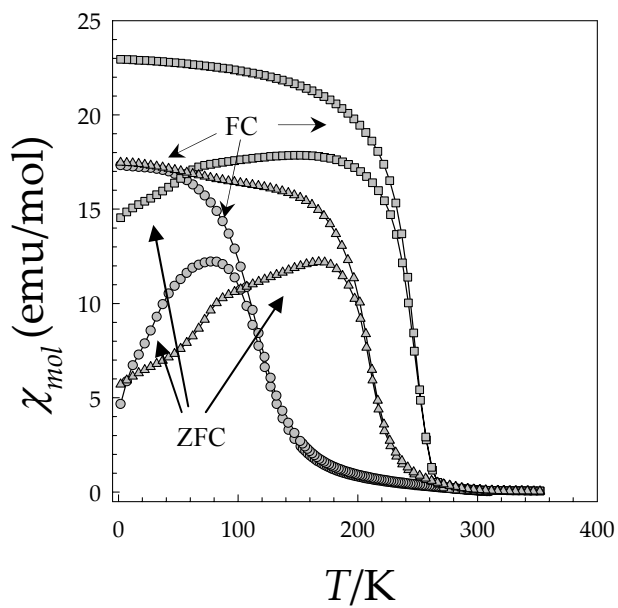

**Fig. 1** ZFC and FC curves for oxygen annealed $La_{1-x}Na_xMn_{1-y}Ru_yO_{3+\delta}$ samples with x = 0.05 and y = 0 (circles), 0.05 (squares) and 0.15 (triangles).

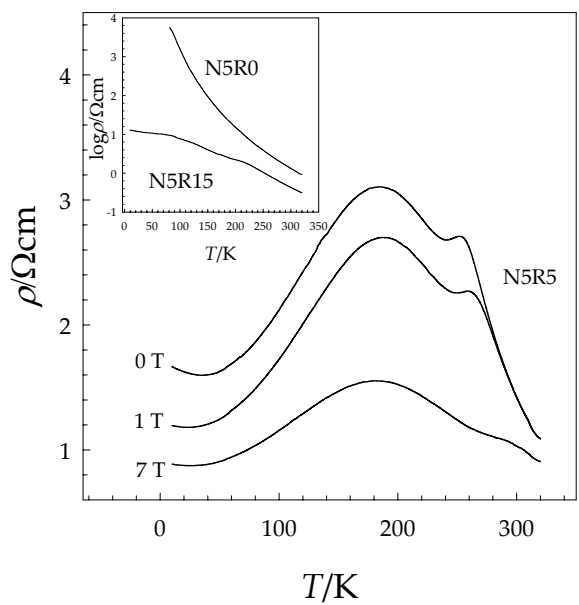

**Fig. 2** Resistivity curves for oxygen annealed N5R5 sample measured at null applied magnetic field (0 T) and at 1 and 7 T. In the inset are reported the 0 T curves for the N5R0 and N5R15 samples on a logarithmic scale.